\pdfoutput=1
\documentclass[aps,prl,twocolumn,showpacs,a4paper]{revtex4}
\usepackage[latin1]{inputenc}
\usepackage{graphicx}
\usepackage{amsmath,amssymb}
\usepackage{hyperref}
\hypersetup{colorlinks=true,
						linkcolor=red,
						anchorcolor=blue,
						citecolor=blue,
						filecolor=blue,
						menucolor=blue,
						urlcolor=blue
}
\usepackage{datetime}
\usepackage{upgreek}
\usepackage{placeins}
\usepackage{units}
\usepackage{xcolor}

\begin{document}

\title{Quantum phases from competing short- and long-range interactions in an optical lattice}

 \author{%
Renate Landig, Lorenz Hruby, Nishant Dogra, Manuele Landini, Rafael Mottl, Tobias Donner and Tilman Esslinger}
 \affiliation{%
Institute for Quantum Electronics, ETH Zurich, 8093 Zurich, Switzerland}
\maketitle
{\bf 
Insights into complex phenomena in quantum matter can be gained from simulation experiments with ultracold atoms, especially in cases where theoretical characterization is challenging. However, these experiments are mostly limited to short-range collisional interactions. Recently observed perturbative effects of long-range interactions were too weak to reach novel quantum phases \cite{Baier2015, Yan2013}.
Here we experimentally realize a bosonic lattice model with competing short- and infinite-range interactions, and observe the appearance of four distinct phases - a superfluid, a supersolid, a Mott insulator and a charge density wave. Our system is based on an atomic quantum gas trapped in an optical lattice inside a high finesse optical cavity. The strength of the short-ranged on-site interactions is controlled by means of the optical lattice depth. The infinite-range interaction potential is mediated by a vacuum mode of the cavity \cite{Baumann2010, Mottl2012} and is independently controlled by tuning the cavity resonance. When probing the phase transition between the Mott insulator and the charge density wave in real-time, we discovered a behaviour characteristic of a first order phase transition. Our measurements have accessed a regime for quantum simulation of many-body systems where the physics is determined by the intricate competition between two different types of interactions and the zero point motion of the particles.
}

Experiments with cold atoms have contributed in many ways to elucidate fundamental behaviour of quantum matter \cite{Bloch2012}. An example is the realization of the Bose-Hubbard model, where the balance between the kinetic energy of particles moving in an optical lattice and the on-site collisional interactions drives a quantum phase transition from a superfluid to a Mott insulating phase \cite{Bloch2002,Kohl2005}. Whilst collisions between atoms are naturally present in quantum gases and give rise to short-range interactions \cite{Weiner1989}, longer ranged interactions are more elusive. In order to get a handle on the latter, ultracold gases of particles with large magnetic or electric dipole moments \cite{Ni2008, Stuhler2005}, atoms in Rydberg states \cite{Heidemann2008}, or cavity-mediated interactions \cite{Baumann2010} have been studied. Indeed, already Hubbard models with additional nearest-neighbour interactions are predicted to show intriguing phases like charge and spin density waves, supersolids, topological phases or checkerboard and stripe phases \cite{Dutta2015, Mickiewicz1990,Goral2002, Kovrizhin2004,VanOtterlo1995,Scarola2005,DallaTorre2006}.

\begin{figure}
\includegraphics[width=0.98\columnwidth]{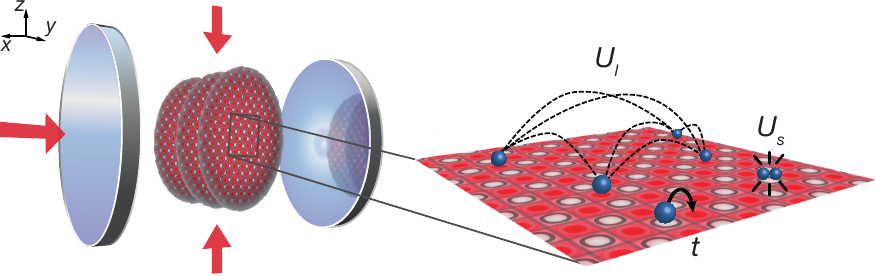}
\caption{
{\bf Illustrations of the experimental scheme realising a lattice model with on-site and infinite-range interactions.} 
A stack of 2D systems along the $y$-axis is loaded into a 2D optical lattice (red arrows). The cavity induces atom-atom interactions of infinite range. Illustration of the competing energy scales: tunnelling $t$, on-site interactions $U_{\mathrm{s}}$ and long-range interactions $U_{\mathrm{l}}$. 
}
\label{fig:BH-scheme}
\end{figure}

In our experiment, we achieve independent control over three energy scales by combining an optical lattice with cavity-mediated interactions, see Fig. 1. The underlying static lattices along all three directions are necessary to study the direct competition between short- and long-range interactions, as compared to the situation very recently investigated in \cite{Klinder2015a}.
Aspects of this scenario, in which on-site interactions compete with infinite-range interactions, have been theoretically studied in the context of self-consistent extended Hubbard models and various phases have been predicted \cite{Ritsch2013, Li2013, Habibian2013a}. 
The starting point is a Bose-Einstein condensate (BEC) of $4.2(4) \times 10^4$ $^{87}$Rb atoms which is prepared inside the ultrahigh-finesse optical cavity.
The optical lattice is formed by three mutually orthogonal standing waves. The lattice along the $y$-axis at wavelength $\lambda_y = 670\, \unit{nm}$ splits the BEC into a stack of about 60 weakly coupled two-dimensional (2D) layers. These 2D layers are then exposed to a square lattice in the $xz$-plane formed by one free space lattice and one intracavity optical standing wave, both at a wavelength of $\lambda = 785.3\, \unit{nm}$. They create periodic optical potentials of equal depths $V_{\mathrm{2D}}$ along both directions, which we will specify in units of the recoil energy $E_{\mathrm{R}}= h^2/2m\lambda^2$, where $m$ denotes the mass of $^{87}$Rb. In addition to the lattice potential, the atoms are exposed to an overall harmonic confinement, which results in a maximum density of 2.8 atoms per lattice site at the center of the trap. The standing wave along the $z$-axis fulfills a second role as it controls long-range interactions via off-resonant scattering into the optical resonator mode. 
The photons are scattered off the trapped atoms and are delocalized within the cavity mode thereby mediating atom-atom interactions of infinite range (see Methods). 
These infinite-range interactions create $\lambda$-periodic atomic density-density correlations on the underlying $\lambda/2$-periodic square lattice \cite{Mottl2012}. The correlations can lead to the breaking of a $\mathbb{Z}_2$-symmetry between the two checkerboard sublattices \cite{Baumann2011}, defined by either even or odd sites, resulting in the appearance of a self-consistent optical potential with alternating strength.

In a wide range of the parameter space, the system can be described by a lattice model with long-range interactions (see Methods and Extended Data Fig. 1), given by:
\begin{eqnarray}
\begin{split}
\hat{H} =  - t \sum_{\langle e,o \rangle}\left(\hat{b}^{\dagger}_{e}\hat{b}_{o} + \mathrm{h}.\mathrm{c}.\right) + \frac{U_{\mathrm{s}}}{2} \sum_{i\in e,o} \hat{n}_{i}(\hat{n}_{i}-1) \\
   - \frac{U_{\mathrm{l}}}{K}\left(\sum_{e}{\hat{n}_{e}} - \sum_{o}{\hat{n}_{o}}\right)^{2} - \sum_{i\in e,o}{\mu_i\hat{n}_{i}}.
\end{split}
\label{eq:BHHam}
\end{eqnarray}
Here $e$ and $o$ denote all even and odd lattice sites respectively, $\hat{b}_i$ $(\hat{b}^{\dagger}_i)$ are the bosonic annihilation (creation) operators at site $i$, $\hat{n}_i$ counts the number of atoms on site $i$ and $K$ is the total number of sites. The first term represents the tunnelling between neighbouring sites at rate $t$ and favours delocalization of the atoms within a 2D layer, supporting superfluidity. The second term describes the on-site interaction with strength $U_{\mathrm{s}}$ controlled via $V_{\mathrm{2D}}$. Its energy is minimized if the atomic wavefunctions are localized on individual lattice sites, with balanced populations on even and odd sites and vanishing spatial coherence. The infinite-range interactions are captured by the third term and favour, for positive $U_{\mathrm{l}}$, a particle imbalance between even and odd sites. This global atom-atom interaction strength $U_{\mathrm{l}}$ is proportional to $V_{\mathrm{2D}}$ and inversely proportional to the detuning $\Delta_{\mathrm{c}} = \omega_z - \omega_{\mathrm{c}}$, where $\omega_z$ is the frequency of the $z$-lattice beam and $\omega_{\mathrm{c}}$ is the cavity resonance frequency (see Methods). The last term represents the effective chemical potential $\mu_i = \mu - \epsilon_{i}$, incorporating the chemical potential $\mu$ and the external trapping potential $\epsilon_{i}$ on lattice site $i$. In the absence of long-range interactions, equation (1) reduces to the Bose-Hubbard model.

To explore the phase diagram of $\hat{H}$, the lattices along the $x$ and $z$-direction are simultaneously ramped up to a certain value $V_{\mathrm{2D}}$, keeping the total ramp time constant. This procedure is repeated for different relative strength of short- and long-range interactions ($U_{\mathrm{l}}$/$U_{\mathrm{s}}$), controlled via the detuning $\Delta_{\mathrm{c}}$. 

In order to detect a superfluid-insulator phase transition, we probe the spatial coherence of the gas by turning off all confining potentials and taking absorption images of the atomic cloud after ballistic expansion. 
Fig. 2a shows measured projected momentum distributions for four different $V_{\mathrm{2D}}$, together with extracted vertical line sums. 
For small lattice depth $V_{\mathrm{2D}}$, spatial coherence can be observed, characterized by a narrow momentum distribution of the cloud and a large BEC fraction $f$, extracted from a bimodal fit to the distribution.
When increasing $V_{\mathrm{2D}}$, the momentum distributions broaden, indicating a drop of coherence and $f$ reduces. 
We observe a kink in $f$ as a function of the interaction strength $U_{\mathrm{s}}/t$ (for details, see Methods and Extended Data Fig. 5), which we associate with the formation of an insulating phase in the cloud and a loss of superfluidity \cite{Jimenez-Garcia2010}. The extracted transition points are shown as white points in Fig. 2b. We confirmed that coherence between different lattice sites is restored when ramping down the 2D lattice potential again. 

\begin{figure*}[]
\includegraphics[width=2.05\columnwidth]{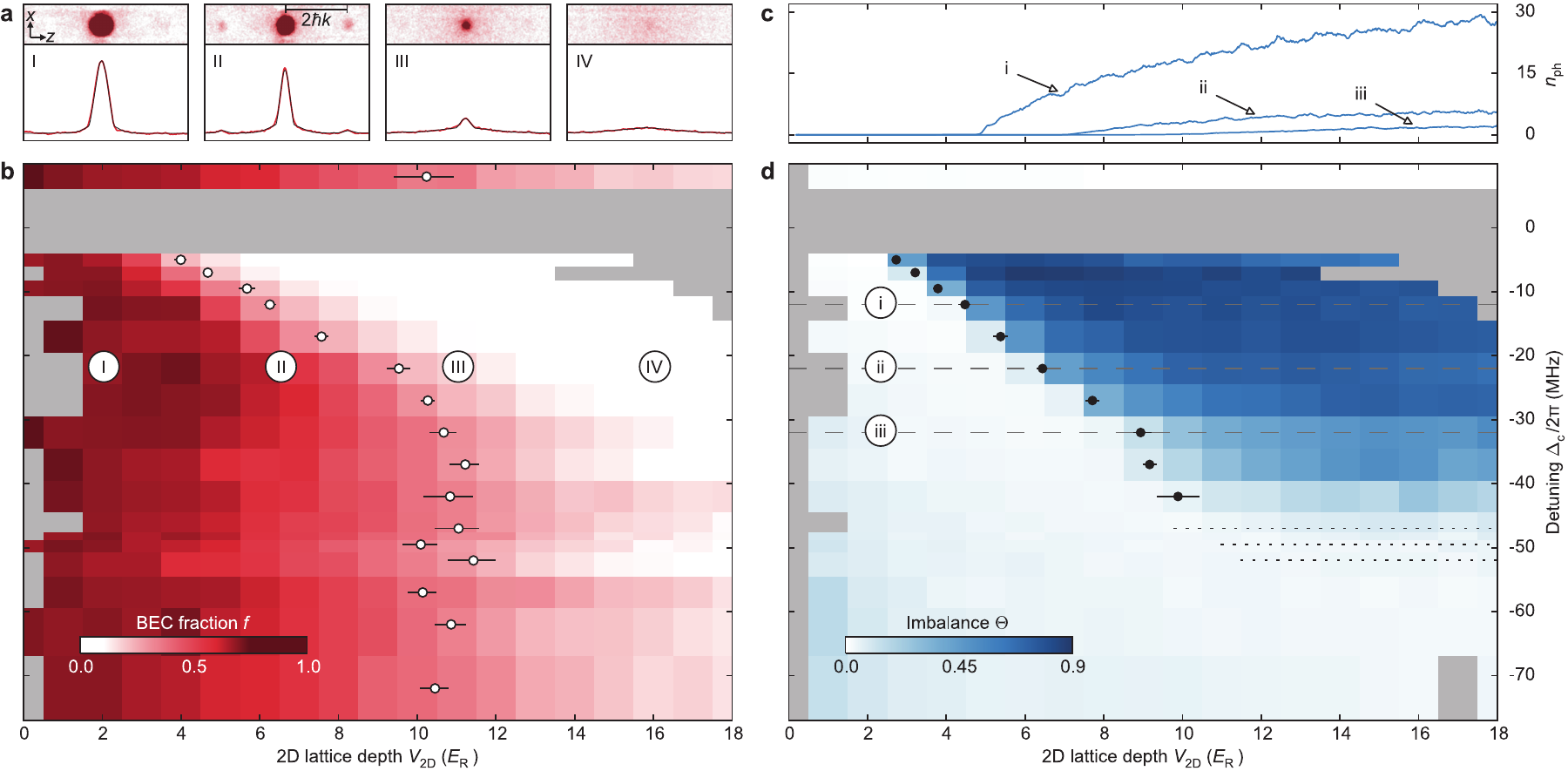}
\caption{
{\bf Characterization of the phases via spatial coherence ({\bf a, b}) and via even-odd imbalance ({\bf c, d}).} 
{\bf a}, Absorption images in the xz-plane (upper panels) and the same signal integrated along the cavity axis (lower panels, red), taken after a ballistic expansion for lattice depths $V_{\mathrm{2D}}$ of $2 \,E_{\mathrm{R}}$ (I), $6.5 \,E_{\mathrm{R}}$ (II), $11\,E_{\mathrm{R}}$ (III) and $16 \,E_{\mathrm{R}}$ (IV) at $\Delta_{\mathrm{c}} /2 \pi =-22\, \unit{MHz}$. Black lines show fits with a bimodal distribution including higher momentum peaks. Due to the cavity mirrors, the field of view along the $x$-direction is restricted.
{\bf b}, Extracted BEC fraction $f$ as a function of $V_{\mathrm{2D}}$ and  $\Delta_{\mathrm{c}}$. White points mark the transition from a superfluid to an insulating phase and are obtained from a piecewise linear fit to the BEC fraction (see Extended Data Fig. 5). Error bars indicate fit uncertainties (see Methods)
{\bf c}, Scattered photons $n_{\mathrm{ph}}$ of single repetitions as a function of $V_{\mathrm{2D}}$ for pump-cavity detunings $\Delta_{\mathrm{c}} /2 \pi$ of $-12\,\unit{MHz}$ (i), $-22\,\unit{MHz}$ (ii) and $-32\,\unit{MHz}$ (iii).
{\bf d}, Imbalance  $\Theta$ mapped as a function of $\Delta_{\mathrm{c}}$ and $V_{\mathrm{2D}}$. We assign the onset of a scattered cavity light field (black points) to the formation of a phase with even-odd imbalance. In the region indicated by the three dotted lines at values $\Delta_{\mathrm{c}}/2 \pi=\{-47, -49.5, -52\} \,\unit{MHz}$, the onset of the cavity light field showed a large variation. Error bars indicate the s.d. of the fit, an additional systematic error of $0.2 \,E_{\mathrm{R}}$ stems from the data analysis. The detection background is growing with decreasing $V_{\mathrm{2D}}$ and increasing detuning from cavity resonance (see Methods). Grey areas were not recorded.} 
\label{fig:observables}
\end{figure*}

An even-odd imbalance causes a $\lambda$-periodic density modulation that acts as a Bragg grating, off which photons from the $z$-lattice beam are scattered into the cavity mode and vice versa. The amplitude of the scattered light field adiabatically follows the atomic density distribution \cite{Mottl2012} and is continuously monitored using a heterodyne detection (see Methods).
Fig. 2c displays mean intracavity photon numbers $n_{\mathrm{ph}}$ measured as a function of $V_{\mathrm{2D}}$. The onset of a cavity field is clearly visible and is taken as the transition point to a phase with even-odd imbalance $\Theta$, marked with black points in Fig. 2d (for details, see Methods and Extended Data Fig. 5). The imbalance $\Theta$ can be quantified using (see Methods):
\begin{eqnarray}
\Theta =  \left|\frac{\sum_{e}{\left\langle\hat{n}_{e}\right\rangle} - \sum_{o}{\left\langle\hat{n}_{o}\right\rangle}}{\sum_{e}{\left\langle\hat{n}_{e}\right\rangle} + \sum_{o}{\left\langle\hat{n}_{o}\right\rangle}}\right|\approx \frac{1}{N}\sqrt{n_{\mathrm{ph}}\, \frac{\Delta_{\mathrm{c}}^2}{\eta^2 }}.  
\label{eq:Theta}
\end{eqnarray}
Here, $\eta$ is the two-photon Rabi frequency of the scattering process and $N$ is the total atom number.

\begin{figure}
\includegraphics[width=0.98\columnwidth]{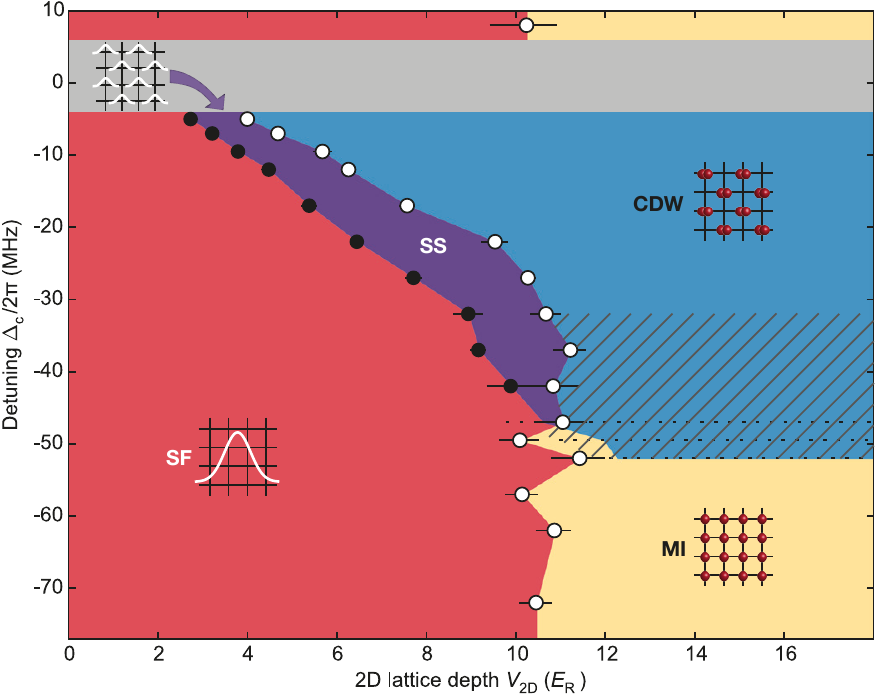}
\caption{%
{\bf Phase diagram.}
{The four phases are indicated by different colors: SF (red), SS (violet), CDW (blue), MI (yellow). Simplified density distributions are schematically illustrated for the homogeneous case with, on average, one atom per site. Data points (from Fig. 2b,d) show the experimentally obtained phase transition points recorded for increasing $V_{\mathrm{2D}}$: Black data points indicate the onset of an even-odd imbalance, white data points depict where spatial coherence is lost. Increasing the 2D lattice depth $V_{\mathrm{2D}}$ simultaneously increases short- and long-range interactions. The detuning $\Delta_{\mathrm{c}}$ changes only the strength of the long-range interactions. The slanted lines indicate the region where CDW and MI may coexist. At detuning $\Delta_{\mathrm{c}} / 2 \pi= +8\, \unit{MHz}$, $U_{\mathrm{l}}$ becomes negative and favours zero imbalance, thus only SF and MI phases appear. No data was taken at detunings indicated by the grey bar. A version of the phase diagram in Hamiltonian parameters is shown in Ext. Data Fig. 2. }
}
\label{fig:phasediagram}
\end{figure}
To establish a phase diagram, we combine all determined transition points in Fig. 3. 
We identify four phases that arise from the competition of the three energy scales: a superfluid (SF), a supersolid (SS), a Mott insulator (MI) and a charge density wave (CDW) phase. 
Far away from cavity resonance, i.e. $\Delta_{\mathrm{c}}/2 \pi\lesssim -52\,\unit{MHz}$, $U_{\mathrm{l}}$ becomes small and the system undergoes, for large enough $V_{\mathrm{2D}}$, a transition from a superfluid to a Mott insulating phase. The latter is characterized by a loss of coherence, as well as the absence of an even-odd imbalance. 
The observed SF to MI transition line is shifted to larger values of $V_{\mathrm{2D}}$ than theoretically expected for a homogeneous system \cite{Krauth2007}, which we attribute to the harmonic confinement of our 2D systems \cite{Rigol2009}.
Approaching cavity resonance increases $U_{\mathrm{l}}$.  Above $\Delta_{\mathrm{c}}/2 \pi\approx -52\,\unit{MHz}$, this leads to the formation of a structured phase with even-odd imbalance, heralded by the onset of a light field scattered into the cavity.
Depending on the relative strength of tunnelling and short-range interactions, the structured phase can either be a SS, where superfluidity is supported, or a CDW phase, where spatial coherence is lost. The identification of the SS phase is further supported by the observation of additional interference peaks corresponding to a $\lambda$-periodic density modulation (see Extended Data Fig. 4) \cite{Kovrizhin2004}.
The SF to SS phase boundary shifts to smaller $V_{\mathrm{2D}}$ when approaching the cavity resonance \cite{Baumann2010}. The transition line from a SS to a CDW follows the same trend. We attribute the loss of coherence in the CDW phase to a reduced nearest-neighbour tunnelling, which has its origin in an energy offset between even and odd lattice sites. This energy offset is a result of the optical potential created by the interference of the field scattered into the cavity with the field of the $z$-lattice, and is shown in Extended Data Fig. 6. Our experimental resolution currently does not allow us to assess the precise topology of the multicritical region.

For long-range interactions dominating over tunnelling and short-range interactions, we observe a maximum even-odd imbalance $\Theta$ above $0.9$, implying that mostly even or odd sites are occupied \cite{Caballero-Benitez2015}. This imbalance is significantly lower between $-32\, \unit{MHz} \gtrsim \Delta_{\mathrm{c}}/2 \pi\gtrsim - 52\,\unit{MHz}$, see Fig. 2d. A possible explanation is the coexistence of CDW and MI phases \cite{Li2013}, which is supported by the external trapping potential, making it energetically costly for the system to place atoms away from the trap center. Contrary, a homogeneous system with non-integer filling would turn into a structured phase with even-odd imbalance for any finite long-range interaction (see Methods and Extended Data Fig. 3). Since the particles in the Mott insulating regions do not scatter into the cavity, the cavity field will rapidly vanish when the size of these regions increases. We conclude from the signal-to-noise ratio of our detection that below $\Delta_{\mathrm{c}}/2\pi = -52\,\unit{MHz}$ the technical noise does not allow us to detect an even-odd imbalance below  $0.01$.

We now study the evolution between the predominantly insulating CDW and MI phases.
We initialize the system in the insulating region ($V_{\mathrm{2D}} = 14\, E_{\mathrm{R}}$) at a certain detuning $\Delta_{\mathrm{c}}$ and then continuously vary $U_{\mathrm{l}}$ by changing $\Delta_{\mathrm{c}}$, before returning to the initial value of $\Delta_{\mathrm{c}}$ (see Methods).
The cavity output field tracks the instantaneous even-odd imbalance $\Theta$ in real-time.
Fig. 4a shows the evolution of the imbalance when decreasing $\Delta_{\mathrm{c}}$ from an initial value in the CDW phase. The data shows a hysteretic behaviour with a lower imbalance on return. The imbalance evolution for a starting value of $\Delta_{\mathrm{c}}$ in the transition region between CDW and MI shows a similar behaviour when decreasing $\Delta_{\mathrm{c}}$ and the opposite behaviour when increasing $\Delta_{\mathrm{c}}$. In Fig. 4c, where we started in the MI phase, the imbalance remains low throughout the measurement. We measured the hysteretic behaviour to be insensitive to the ramp speed, see Extended Data Fig. 7.
\begin{figure}
\includegraphics[width=0.98\columnwidth]{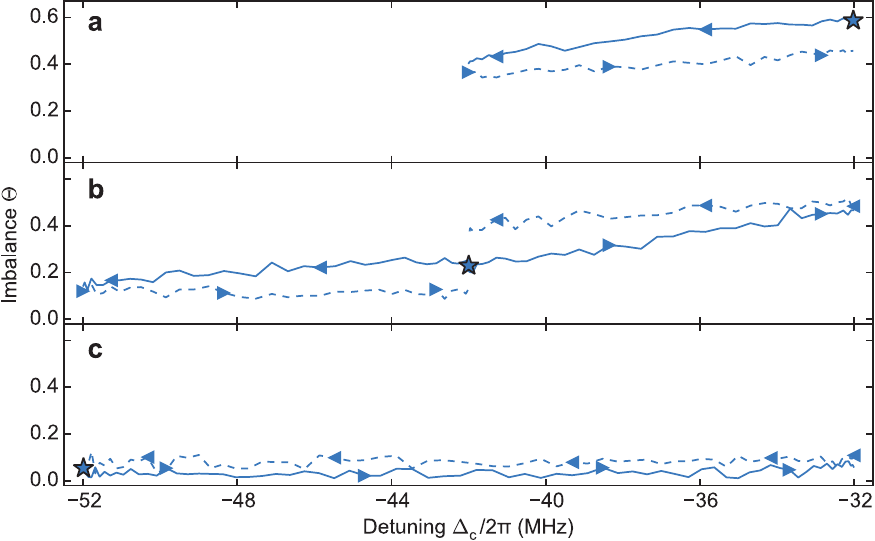}
\caption{%
{\bf Hysteretic behaviour of the CDW to MI transition.} 
 Imbalance $\Theta$ recorded by varying $\Delta_{\mathrm{c}}$ at a rate of $0.67 \,\unit{MHz/ms}$, for fixed $V_{\mathrm{2D}} = 14\, E_{\mathrm{R}}$. The initial detunings $\Delta_{\mathrm{c}}/2 \pi$, indicated by stars, are $-32 \,\unit{MHz}$ ({\bf a}), $-42 \,\unit{MHz}$ ({\bf b}) and $-52\, \unit{MHz}$ ({\bf c}). Arrows signify the ramp directions; dashed lines show the return to the starting point. Curves are rescaled to take atom loss into account and contain three to nine averages, binned at $400\,\unit{\mu s} $ (see Methods).
}
\label{fig:freqscan}
\end{figure}
This hysteretic behaviour of the system points towards a first order phase transition between CDW and MI. When starting in a MI and increasing $U_{\mathrm{l}}/U_{\mathrm{s}}$ beyond a certain point, the CDW will become energetically favourable, but cannot be reached due to an energy barrier between the two phases. Further increasing $U_{\mathrm{l}}/U_{\mathrm{s}}$ lowers this energy barrier until the system is driven out of the metastable state. We suggest that this is activated in our inhomogeneous system by $\lambda$-periodic density-density correlations that are created in residual compressible regions, or superfluid shells, acting as impurities. In the opposite direction, moving from a CDW to a MI phase, the energy offset between even and odd lattice sites stabilizes the CDW phase beyond the point where the MI gets energetically favourable, which results in the observed hysteretic behaviour.  

\section*{METHODS}

\setcounter{figure}{0}
\renewcommand{\figurename}{Extended Data Figure}
\renewcommand{\tablename}{Extended Data Table}

\subsection{Preparation of a BEC in 2D layers.}
We produce a Bose-Einstein condensate (BEC) of $4.2(4) \times 10^4$ $^{87}$Rb atoms at a temperature of $42(2) \,\unit{nK}$ in the $|F,m_F\rangle = |1,-1\rangle$ hyperfine state where $F$ and $m_{F}$ are the total angular momentum and the corresponding magnetic quantum number. The quantization axis is defined by a magnetic field pointing along the $z$-direction. 
The BEC is confined to the center of a TEM$_{00}$ mode of the cavity by an optical dipole trap at a wavelength of $852 \,\unit{nm}$, with trap frequencies of $\omega_{x,y,z}/2 \pi = \left[70.6(3),\,31.4(5),\,29.4(2)\right] \,\unit{Hz}$. Further details on the cavity setup can be found in \cite{Baumann2010}.

The trapped BEC is loaded into a blue-detuned optical lattice of wavelength $\lambda_y=670 \,\unit{nm}$ oriented along the $y$-direction. This is done by implementing a smooth amplitude ramp (S-ramp) in time $t$ which is of the form: $V(t) = V_0\left[3\left(t/t_0\right)^2 - 2\left(t/t_0\right)^3 \right]$, where $V_0$ is the final lattice depth and $t_0$ is the total duration of the ramp. The lattice depth is increased to a final value of $24.9(1)\,E^{670}_{\mathrm{R}}$ in $100 \,\unit{ms}$, where ${E^{670}_{\mathrm{R}}}={{h}^2}/{2{m}{\lambda^2_y}}$ is the atomic recoil energy with $m$ being the mass of a $^{87}$Rb atom. The trap frequencies are kept constant during the loading by increasing the dipole trap depth simultaneously with the blue-detuned lattice. In this way, the whole BEC is cut into roughly 60 2D layers with about 1300 atoms in the central layer.

\subsection{Loading into the square lattice.}
After the preparation of 2D layers, the BEC is exposed to a 2D optical lattice in the $xz$-plane at a wavelength of $\lambda=785.3\,\unit{nm}$. The lattice along the $z$-direction is formed by a free space retro-reflected standing wave laser field which is linearly polarized along the $y$-direction. The lattice along the $x$-direction is created by pumping the TEM$_{00}$ mode of the cavity with linear polarization along the $z$-direction. The effect of interference between the $x$- and $z$-lattices on atoms is minimized by introducing a frequency offset of at least $5\, \unit{MHz}$ between the two laser frequencies. Both lattices are ramped simultaneously within a fixed time of 50 $\unit{ms}$ to a variable lattice depth $V_{\rm{2D}}$ using again the S-ramp. The lattice potential seen by the atoms is of the form: $V(x,z) = V_{\mathrm{2D}}\left[\cos^2(kx) + \cos^2(kz)\right]$, where $k=2 \pi /\lambda$ is the wave number and $V_{\mathrm{2D}}$ is the depth of the lattice in units of the corresponding recoil energy $E_{\mathrm{R}}=h^2/2 m\lambda^2$.

\subsection{Characterization of the optical lattices.}
The lattice depths along the $y$- and $z$-direction are calibrated using Raman-Nath diffraction \cite{Morsch2006}, whereas the lattice depth along the $x$-direction is calibrated using amplitude modulation spectroscopy between the lowest and the first two excited Bloch bands \cite{Stoferle2004}. We estimate the calibration uncertainties on all lattice depths to be smaller than $4\%$. The uncertainty in the intracavity optical lattice depth is enlarged to about $10\%$ by shifts of the cavity resonance frequency due to atomic redistribution during the $V_{\rm{2D}}$ ramp and residual drifts of the incoupling to the resonator.

The heating effect of the near-resonant $xz$-optical lattices on the BEC is characterized by ramping back down the lattices after reaching the insulating regime. We recover a BEC fraction larger than 0.45 and observe an atom loss of 5$-$10$\%$. Loading of lattices in the $xz$-plane also increases the overall confinement. The trap frequencies are $\omega'_{x,z}/2 \pi  = \left[170,\,165\right]\,\unit{Hz}$ at a typical lattice depth of $10\,E_{\mathrm{R}}$.

\subsection{Lattice model with long-range interactions.}
The single-particle Hamiltonian $\hat{H}_{\rm{sp}}$, describing the dynamics of an atom strongly coupled to a single cavity mode and moving in a 2D layer in the presence of static optical lattices, is given as \cite{Baumann2010, Ritsch2013}:
\begin{eqnarray}
\begin{split}
\hat{H}_{\rm{sp}} &= \hat{H}_0 + {V}_{\mathrm{Trap}}(x,z) + \hbar \eta (\hat{a}^{\dagger} + \hat{a}) \cos (kx)\cos (kz) \\
& - \hbar \left(\Delta _{\mathrm{c}} - {U}_{0} \cos ^{2}(kx)\right)\hat{a}^{\dagger}\hat{a}.
\end{split}
\end{eqnarray}
$\hat{H}_0$ consists of the kinetic energy of the particle and the potential seen due to the optical lattices in the $xz$-plane:
\begin{equation}
\hat{H}_0 = \frac{\hat{p}^{2}_x}{2m} + \frac{\hat{p}^{2}_z}{2m} + V_{\mathrm{2D}} \left(\cos ^{2} (kx) + \cos ^{2} (kz) \right).
\end{equation}
$V_{\rm{Trap}}(\textit{x},\textit{z})$ incorporates the inhomogeneous confining potential seen by the atoms. $\hat{a}$ ($\hat{a}^{\dagger}$) annihilates (creates) a photon in the cavity mode. Scattering of the light field from the $z$-lattice into the cavity mode at a two-photon Rabi frequency $\eta$ creates a self-consistent checkerboard lattice for the atoms and is represented by the third term in $\hat{H}_{\rm{sp}}$. This term describes how the atomic motion self-consistently determines the occupation of the cavity field mode inducing infinite-range interactions between the atoms. The last term in $\hat{H}_{\rm{sp}}$ represents the cavity field in the rotating frame of the $z$-lattice with $\Delta_{\mathrm{c}} = \omega_z - \omega_{\mathrm{c}}$. The effect of the dispersive shift of the cavity resonance frequency is also included with $U_0$ being the maximum light shift per atom.
 
The many-body description of the system is obtained by introducing the bosonic field operator $\hat{\Psi}({\bf r})$ ($\hat{\Psi}^{\dagger}(\textbf{r})$) which annihilates (creates) a particle at position $\textbf{r} = (x,z)$ and satisfies bosonic commutation relations. In the framework of second quantization, the many-body Hamiltonian $\hat{{H}}^{\mathrm{2nd}}$ reads:
\begin{equation}
\begin{split}
\begin{aligned}
\hat{{H}}^{\mathrm{2nd}} &= \int \mathrm{d}\textbf{r}\,\hat{\Psi}^{\dagger} (\textbf{r}) \Big[\hat{H}_{\mathrm{sp}} - \mu \\ & \mathrel{\phantom{=\int}}\quad{} +g_{\rm{2D}}\hat{\Psi}^{\dagger}(\textbf{r})\hat{\Psi}(\textbf{r})\Big]\hat{\Psi} (\textbf{r}),
\end{aligned}
\end{split}
\end{equation}
where $\mu$ is the chemical potential and $g_{\rm{2D}}$ is the modified short-range interaction strength in a 2D layer \cite{Petrov2000}. We expand $\hat{\Psi}(x,z)$ in the basis of Wannier functions localized on different lattice sites which are obtained from the lowest Bloch band defined by $\hat{H}_{\rm{0}}$:
\begin{equation}
\hat{\Psi}=\sum_{\textbf{m}} W_{\textbf{m}}(x,z)\hat{b}_{\textbf{m}},
\end{equation}
where $\hat{b}_{\textbf{m}}$ ($\hat{b}^{\dagger}_{\textbf{m}}$) represent the annihilation (creation) operators of a single particle at site $\textbf{m}=(m_x,m_z)\frac{\lambda}{2}$ and $W_{\textbf{m}}(x,z)$ is the Wannier function localized on site $\textbf{m}$. A site is referred to as even (odd) if $m_x+m_z$ is even (odd). The Wannier functions localized on neighbouring  lattice sites are related to each other by a translation of the lattice constant. Keeping interactions only up to the nearest neighbouring sites, we obtain the Bose-Hubbard model with additional terms \cite{Maschler2007}:
\begin{eqnarray}
\begin{split}
\hat{H}^{\mathrm{2nd}}_{\rm{Wan}} &=
- t \sum_{\langle e,o \rangle}\left(\hat{b}^{\dagger}_{e}\hat{b}_{o} + \mathrm{h}.\mathrm{c}.\right) + \frac{U_{\mathrm{s}}}{2} \sum_{i\in e,o} \hat{n}_{i}\left(\hat{n}_{i}-1\right) \\
&+ \hbar\eta M_0\left(\hat{a}^{\dagger} + \hat{a}\right)\left(\sum_{e}{\hat{n}_{e}} - \sum_{o}{\hat{n}_{o}}\right) \\
&- \hbar \left(\Delta_{\rm{c}}-\delta\right) \hat{a}^{\dagger}\hat{a} - \sum_{i\in e,o}{\mu_i\hat{n}_{i}},
\end{split}
\label{eq:semi_BHHam}
\end{eqnarray}
where $t$ and $U_{\mathrm{s}}$ represent tunnelling and contact interaction in the Bose-Hubbard model \cite{Jaksch1998} and are defined as:
\begin{equation}
t = \int\int \mathrm{d}x \,\mathrm{d}z \,W_i^{*}(x,z) \,\hat{H}_0 \,W_i(x,z-\lambda/2)
\label{eq:t}
\end{equation}
\begin{equation}
U_{\rm{s}} = g_{\rm{2D}}\int\int \mathrm{d}x \, \mathrm{d}z\, |W_i(x,z)|^4,
\label{eq:Us}
\end{equation}
$\mu_i = \mu-\epsilon_i$ describes the local chemical potential at site $i$ incorporating the effect of $V_{\mathrm{Trap}}$ and $\hat{n}_{i} = \hat{b}^{\dagger}_i\hat{b}_i$ counts the number of particles on site $i$. Indices $e$ and $o$ refer to all even and odd lattice sites, respectively. $\delta = U_{\rm{0}}M_{\rm{1}}N$ is the dispersive shift of the cavity due to the BEC with $N$ being the total number of atoms. The two overlap integrals $M_{\rm{0}}$, $M_{\rm{1}}$ are defined as:
\begin{eqnarray}
M_{\rm{0}} &=& \int\int \mathrm{d}x \,\mathrm{d}z \,W_i^{*}(x,z) \cos{\left(kx\right)} \cos{\left(kz\right)} W_i(x,z) \notag \\
M_{\rm{1}} &=& \int\int \mathrm{d}x \, \mathrm{d}z\, W_i^{*}(x,z)\cos^2{\left(kx\right)} W_i(x,z) \notag .
\end{eqnarray}
 A higher order correction to the tunnelling along the $x$-direction by the self-consistent cavity lattice is neglected. The cavity decay rate $\kappa$ is large compared to the atomic recoil frequency which allows us to adiabatically eliminate the cavity field \cite{Baumann2010}. Its steady state value is given by:
\begin{equation}
\hat{a} = \frac{ \eta M_{0} }{ \Delta_{c}-\delta + i \kappa }\left(\sum_{e}{\hat{n}_{e}} - \sum_{o}{\hat{n}_{o}}\right).
\label{eq:cavity_field}
\end{equation}
Hence the light leaking out of the cavity will be proportional to the imbalance of the number of atoms on the two kind of sites and it can herald the presence of a phase with broken $\mathbb{Z}_2$-symmetry of the underlying static lattice. Inserting equation (10) into equation (7), we recover equation (1) of the main text, with cavity-mediated long-range interaction strength $U_{\rm{l}}$ given by:
\begin{eqnarray}
\begin{split}
&U_{\rm{l}} = -K \hbar |\eta M_{0}|^{2} \frac{\Delta_{\rm{c}}-\delta}{(\Delta_{\rm{c}}-\delta)^{2} + \kappa^{2}} \\
&\hspace{-0.3cm}\stackrel{\text{$|\Delta_{\rm{c}}| \gg \kappa, |\delta |$}}{\approx} -K \hbar |M_{0}|^{2} \frac{\eta^2}{\Delta_{\rm{c}}} \propto \frac{V_{\mathrm{2D}}}{\Delta_{\rm{c}}}.
\end{split}
\label{eq:Ul}
\end{eqnarray}
 
To describe our stack of 2D layers within this theoretical framework, we assume that a system of many 2D layers can be combined to form one 2D layer with accordingly larger number of lattice sites, containing all atoms. 

\textit{Validity of the theoretical model}: In the derivation of equation (1), we assume the validity of the single-band approximation. To deduce the experimental parameter space where this assumption holds, we compare the strength of all Hamiltonian parameters with the excitation energy to the next higher Bloch band, as shown in Extended Data Fig. 1.
\begin{figure}
\includegraphics[width=0.98\columnwidth]{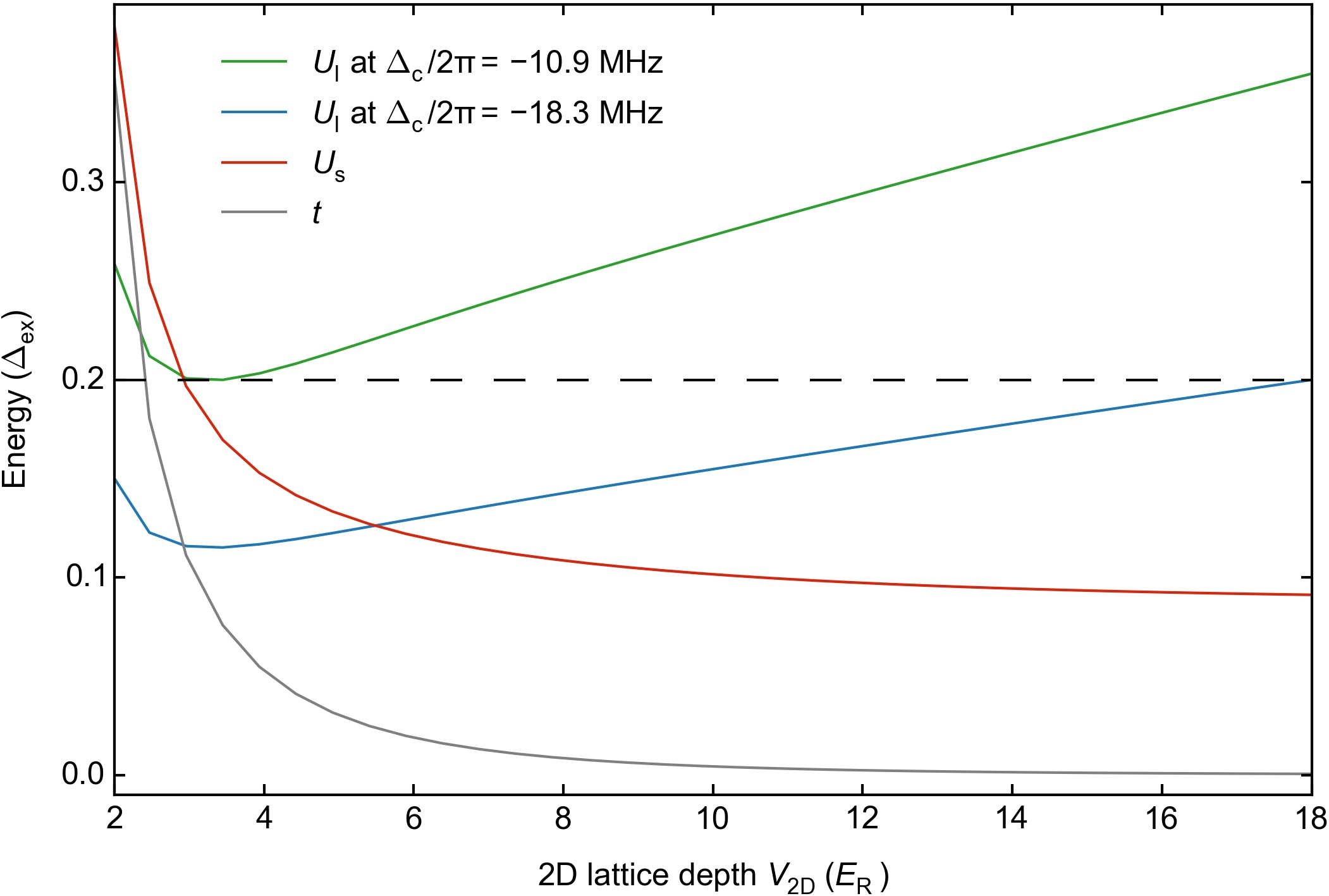}
\caption{{\bf Validity of the single-band approximation.} The energy scales of the Hamiltonian are plotted in units of the minimum gap $\Delta_{\rm{ex}}$ between the lowest and the first excited Bloch band. The single-band approximation is assumed to be valid if all the energy scales $U_{\rm{s}}$, $U_{\rm{l}}$ and $t$ are at least 5 times smaller than $\Delta_{\rm{ex}}$, i.e. if they lie below the black dashed line. This criterion is fulfilled for $\Delta_{\rm{c}}/2\pi < -18.3\,\unit{MHz}$ and $18\,E_{\rm{R}}>V_{\rm{2D}} > 3\,E_{\rm{R}}$. For detunings in the interval $-18.3 \,\unit{MHz} < \Delta_{\rm{c}}/2\pi < -10.9\,\unit{MHz}$, the approximation is only partially valid, depending on $V_{\rm{2D}}$. We use this information to illustrate the region of validity in Extended Data Fig. 2.}
\label{fig:extsingle_band}
\end{figure}
\subsection{Phase diagram in terms of Hamiltonian parameters.}
\begin{figure}
\includegraphics[width=0.98\columnwidth]{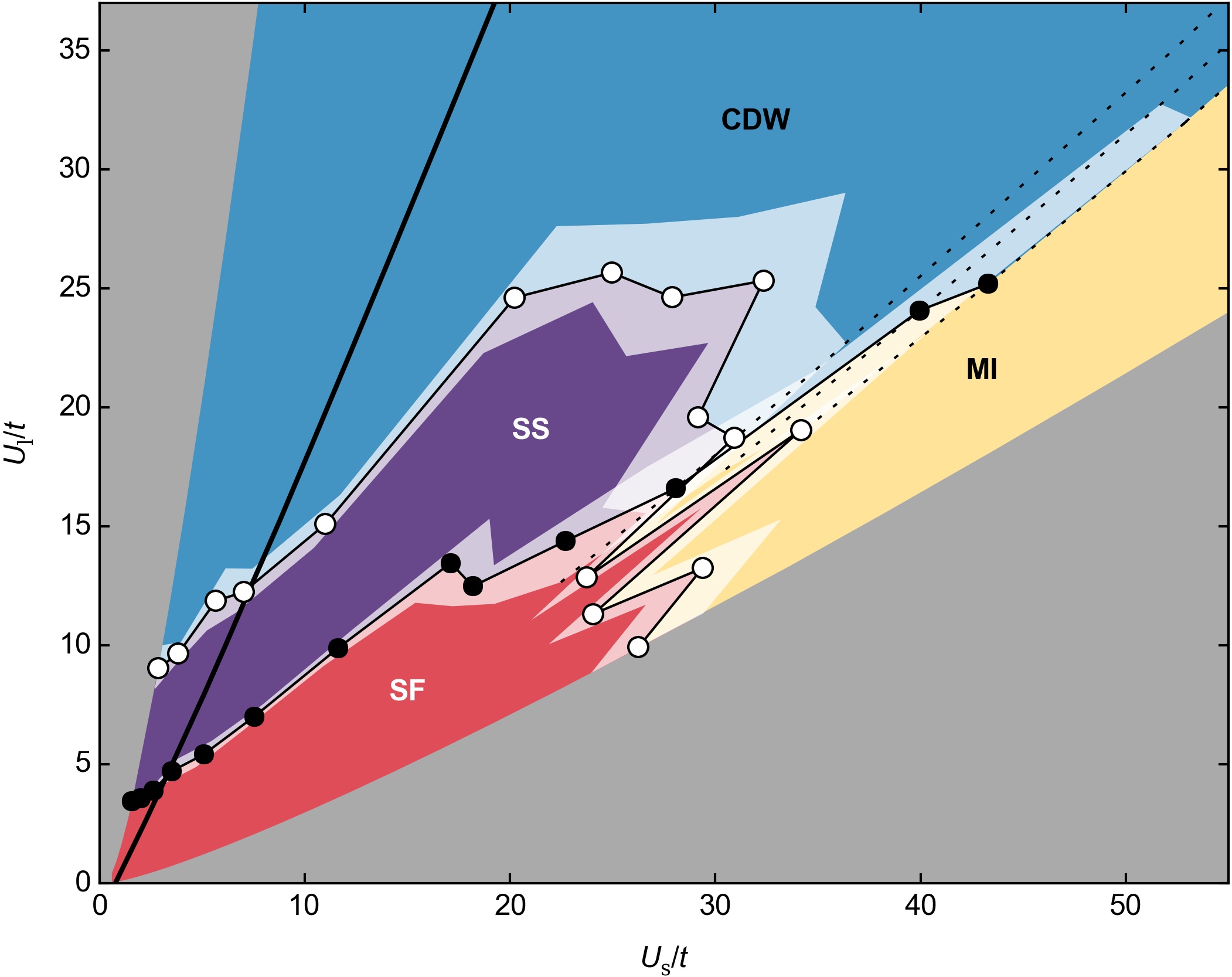}
\caption{{\bf Phase diagram plotted as a function of Hamiltonian parameters $U_{\rm{s}}/t$ and $U_{\rm{l}}/t$.} These are derived by converting experimental parameters from Fig. 3 into Hamiltonian parameters. The region of validity for this conversion lies to the right of the solid black line, grey areas were not recorded. The white data points indicate where spatial coherence is lost, and the black data points depict the onset of an even-odd imbalance. The white shaded regions around the data points represent the respective converted error bars. The dotted black lines show, as in Fig. 3, the region where the onset of the cavity light field showed a large variation.}
\label{fig:extphase_diagram}
\end{figure}
We convert the phase diagram from Fig. 3 into Hamiltonian parameters (see Extended Data Fig. 2) using equations (8), (9) and (11), taking into account the effect of two nearly-degenerate polarization modes of the cavity in the definition of $U_{\rm{l}}$. 
Starting in a SF phase and increasing $U_{\rm{l}}/t$ takes the system into a SS phase and eventually into a CDW phase. At the transition from the SF to SS phase, the system needs to overcome additional short-range interaction energy. As a result, an increasingly larger critical long-range interaction strength is required to enter the SS phase for increasing $U_{\rm{s}}/t$. A similar effect is seen for the transition from a SS to a CDW phase.

For negligible tunnelling, a direct transition from a MI to a CDW phase is found at a relative strength of $\frac{U_{\rm{l}}}{U_{\rm{s}}}= 0.66(4)$. In the absence of tunnelling and trapping potential the Hamiltonian (1) supports a stable MI only for commensurate filling. In this case, the phase boundary between MI and CDW lies at a relative strength of $\frac{U_{\rm{l}}}{U_{\rm{s}}} = 0.5$. Deviations from this value can be attributed to the presence of the trap, incommensurate filling and the non-local nature of the long-range interactions.

For negligible $U_{\mathrm{l}}$, the transition from SF to MI is observed at a relative strength of $\frac{U_{\rm{s}}}{t} = 28(4)$. The value is larger than the theoretically predicted value of $\frac{U_{\rm{s}}}{t} \approx 16$ for a homogeneous system with unity filling \cite{Krauth2007}, as discussed in the main text.

\subsection{Effect of the trapping potential}
The harmonic trapping potential experienced by the atoms has a stabilizing effect on the MI phase in the presence of long-range interactions. Extended Data Fig. 3 illustrates, for fixed atom number and zero tunnelling, the effect of the trapping potential in the presence of the self-consistent lattice potential. For any non-zero $U_{\rm{l}}$ and for non-integer filling, the homogeneous system will arrange in a structured phase with even-odd imbalance. However, the presence of a trapping potential can favour the coexistence of MI and CDW phases or of a MI phase alone, since the system has to pay additional energy for arranging atoms away from the trap center. This energy cost has to be compared with the gain in energy due to the formation of a CDW phase.
For larger fillings, we expect that the system develops a wedding cake like structure similar to experimental realizations of MI phases. In our system, the plateaus can also host partially modulated and fully modulated CDW phases. The presence of any CDW in the system will be signalled by a finite light field scattered into the cavity. We do not expect a qualitative change of the phase diagram depending on the steepness of the trap.
\begin{figure}
\includegraphics[width=0.98\columnwidth]{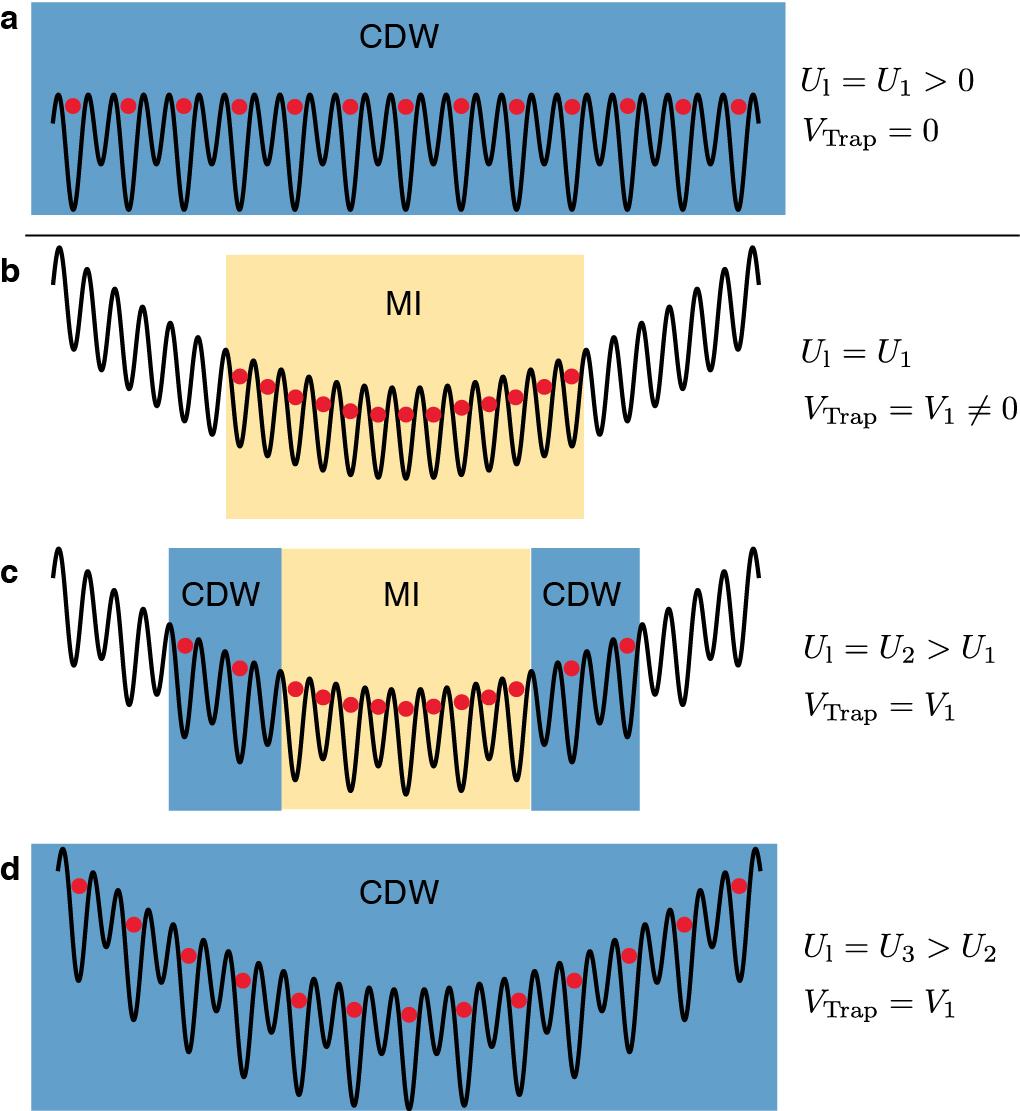}
\caption{{\bf Influence of the trapping potential on the long-range interacting system.} Shown are sketches of a 1D slice through a 2D layer, displaying the ground state configurations of 13 particles depending on the relative influence of trapping potential and long-range interaction. Panel (a) shows the situation for a homogeneous system with finite $U_{\rm{l}}$. Panels (b)-(d) show the state of the system for increasing $U_{\rm{l}}$, starting with small but finite $U_{\rm{l}}$.}
\label{fig:ext1a}
\end{figure}

\subsection{Extraction of the BEC fraction.}
We take an absorption image of the atomic distribution in the $xz$-plane after $15 \,\unit{ms}$ of ballistic expansion. The obtained momentum distribution is integrated over the cavity direction. We perform a bimodal fit to the resulting distribution, in which we distinguish two contributions. The first component represents coherent atoms diffracted by the lattice potential, captured by a Thomas-Fermi profile plus two Gaussian interference peaks at $\pm 2\hbar k$. The second component is a broad Gaussian distribution resulting from the incoherent addition of atomic signals from the insulating part of the cloud. The BEC fraction $f$ is finally extracted from the ratio $f=N_\mathrm{c}/N$, where $N_\mathrm{c}$ is the integrated atom number in the coherent part and $N$ is the total atom number. 
$N$ is obtained from the mean total atom number of all experimental data at low lattice depths, $V_{\mathrm{2D}}\leq 2 \,E_{\mathrm{R}}$. For deeper lattices, the growing spatial extent of the incoherent background is affected by inhomogeneities in the imaging and by the cropped field of view.

To constrain the number of free fit parameters, we fix the position of the interference peaks with respect to the central peak. Furthermore, their widths are linearly correlated to the width of the central peak \cite{Spielman2008}, see Fig. 2a. We double count the interference peaks to correct for the non-visible peaks along the $x$-direction, where the field of view is cropped by the cavity mirrors. The contribution from the $\pm2 \hbar k$ peaks to the total atom number is on the order of a few percent at most. The checkerboard lattice in the supersolid phase leads to extra interference peaks, which lie outside the field of view (see Extended Data Fig. 4). Their contribution to the overall atom number is even lower than the one from the $\pm2 \hbar k$ peaks in most parts of the phase diagram and is therefore neglected.
\begin{figure}
\includegraphics[width=50mm]{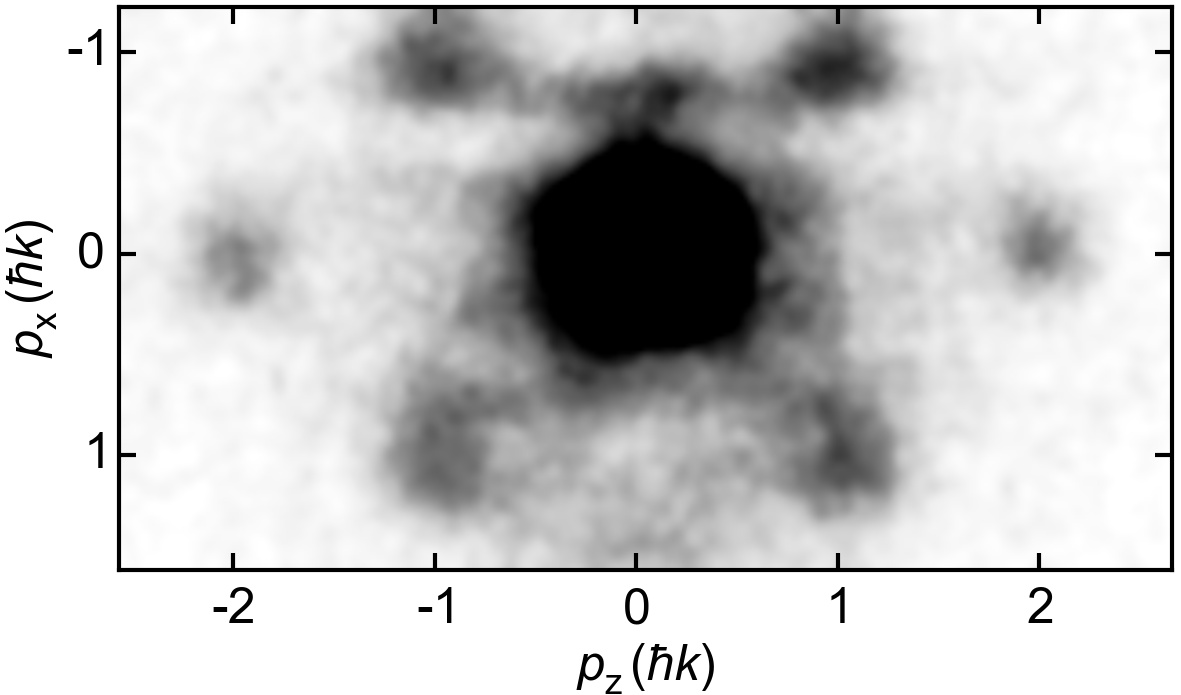}
\caption{{\bf Momentum distribution in the SS phase.} Absorption image from a calibration measurement taken after a short ballistic expansion of $7\,\unit{ms}$ at a detuning of $\Delta_\mathrm{c}/2\pi=-23\,\unit{MHz}$ and a lattice depth $V_{\mathrm{2D}}$~39\% above the onset of an even-odd imbalance in the SS phase. We observe interference peaks at $p_z =\pm 2\hbar k$. Additional interference peaks resulting from the emerging checkerboard lattice appear at $(p_x,p_z) = (\pm hk, \pm hk)$. This observation indicates a SS phase. These additional momentum peaks lie outside the field of view for the longer ballistic expansion time of $15\,\unit{ms}$.} 
\label{fig:ext4}
\end{figure}
\subsection{Extraction of the even-odd imbalance.}
During each experimental sequence, the Bragg scattered light leaking out of the cavity is detected with a heterodyne setup \cite{Landig2015} having a sensitivity of $0.67(1) \,\unit{V}^2$ per intracavity photon. The heterodyne detection is insensitive to the laser field creating the static lattice along the cavity axis by the choice of orthogonal polarizations and a minimum frequency difference of $5 \,\unit{MHz}$. Both phase and magnitude of the light field are recorded. To separate out the coherent part of the light field, we apply a low pass filter to the quadratures before taking the absolute square to obtain an intracavity photon number $n_\mathrm{ph}$.
It is mapped to an even-odd particle imbalance obtained from equation~(10) with $n_{\mathrm{ph}} = \langle \hat{a}^{\dagger}\hat{a}\rangle$. We define the effective even-odd imbalance $\Theta$ under the assumption of completely localized atoms on either even or odd sites ($M_0 = 1$):
\begin{eqnarray}
\Theta = \left| \frac{\sum_{e}{\left\langle\hat{n}_{e}\right\rangle} - \sum_{o}{\left\langle\hat{n}_{o}\right\rangle}}{\sum_{e}{\left\langle\hat{n}_{e}\right\rangle} + \sum_{o}{\left\langle\hat{n}_{o}\right\rangle}}\right| = \frac{1}{N}\sqrt{n_{\mathrm{ph}}\, \frac{\Delta_{\mathrm{c}}^2}{\eta^2 }}\frac{1}{F(\Delta_{\mathrm{c}})},
\label{eq:Theta}
\end{eqnarray}
with 
\begin{eqnarray}
\begin{split}
F&(\Delta_{\mathrm{c}})= \sqrt{\frac{\Delta_{\mathrm{c}}^2\cos^2(\alpha)}{(\Delta_{\mathrm{c}}-\delta-\frac{\delta_{\mathrm{B}}}{2})^2+\kappa^2} + \frac{\Delta_{\mathrm{c}}^2 \sin^2(\alpha)}{(\Delta_{\mathrm{c}}-\delta+\frac{\delta_{\mathrm{B}}}{2})^2+\kappa^2}}\notag\\
&\hspace{-0.12cm}\stackrel{\text{$|\Delta_{\rm{c}}| \gg \kappa, |\delta |,\delta_{\mathrm{B}}$}}{\approx} 1 \notag
\end{split}
\end{eqnarray}
describing the scattering into two linearly polarized $\mathrm{TEM}_{00}$ eigenmodes of the cavity, separated due to birefringence by $\delta_{\mathrm{B}} = 2\pi\times 2.2\,\unit{MHz}$ and oriented at an angle $\alpha=22^\circ$ with respect to the $y$- respectively $z$-axis.
The cavity decay rate $\kappa$ is $2\pi\times 1.25\,\unit{MHz}$ and the effective two-photon Rabi frequency $\eta$ for scattering into the two cavity modes is given by $\eta = 2\pi \times 2.7\sqrt{V_{\mathrm{2D}}/\hbar}\,\sqrt{\mathrm{Hz}}$. Close to cavity resonance, the polarization of the Bragg scattered cavity field rotates slightly due to the birefringence, which we include in the detection efficiency of the heterodyne detection. The maximum dispersive shift ${U}_{0}$ per atom of each of the two cavity modes is $-2\pi\times 45.9 \,\unit{Hz}$. The intracavity photon number is determined with a systematic uncertainty of $8\%$, leading to a relative uncertainty in $\Theta$ of less than $6\%$. The technical background level of the photodetection is converted into an imbalance background, which depends on $\Delta_{\rm{c}}$ and $V_{\mathrm{2D}}$. This causes the signal in the lower left corner of Fig. 2d. However, in the MI phase, we estimate from the s.d. of this background a resolution for $\Theta$ which is better than $1\%$.

\subsection{Phase boundaries.}
\textit{Coherence:} We convert the 2D lattice depth $V_{\mathrm{2D}}$ to the corresponding ratio $U_{\mathrm{s}}/t$ of short-range interaction strength and nearest neighbour tunnelling by using the Wannier functions obtained from the lowest Bloch band of the applied static lattices. In this way, we obtain a BEC fraction $f$ as a function of $U_{\mathrm{s}}/t$ (see Extended Data Fig. 5), which we fit with a piecewise linear function. The first kink in the fit is associated with the transition point to an insulating phase \cite{Jimenez-Garcia2010}. By analyzing the stability of the fit with respect to initial parameters, we deduce an additional uncertainty on top of the s.d. and include it in the error bar displayed for each transition point in Fig. 2b and 3.

\textit{Even-odd imbalance:} From each experimental repetition, we obtain a time trace of the light field scattered into the cavity. The maximum photon number $n_{\mathrm{ph,max}}$ at the end of the trace and the corresponding lattice depth $V_{\mathrm{2D}}$ are averaged in a time window of 10~ms (spanning 7.8\% of $V_{\mathrm{2D}}$), resulting in one data point extracted per time trace. For each detuning, $n_{\mathrm{ph,max}}(V_{\mathrm{2D}})$ is fitted with a piecewise linear and power law function (see Extended Data Fig. 5) to determine the point where the intracavity light field starts building up. This method largely increases the signal to noise ratio while keeping systematic shifts of the onset point to below 0.2~$E_{\mathrm{R}}$. In the region of $-52\,\mathrm{MHz}\leq\Delta_\mathrm{c}/2\pi\leq-47\,\mathrm{MHz}$, the intracavity field becomes very small and fluctuates strongly from shot to shot (see Extended Data Fig. 5c). We therefore indicate a region for the transition to a phase with $\lambda$-periodic density modulation by dashed lines in Fig. 2d and 3. The starting point of these dashed lines indicate the earliest onset of an even-odd imbalance including the s.d. of the fit. Due to the fixed time of the lattice ramp, we cross the transition to an even-odd imbalanced phase non-adiabatically when ramping into deep lattices. The non-adiabaticity leads to a small shift of the onset point towards higher lattice depths, which can be seen in Fig. 2c. This behaviour was studied previously and explained with Kibble-Zurek theory \cite{Baumann2011}. The described method of discretizing the data is intrinsically less sensitive to this type of shift compared to fitting a single time trace to extract the transition point.
\begin{figure}
\includegraphics[width=0.98\columnwidth]{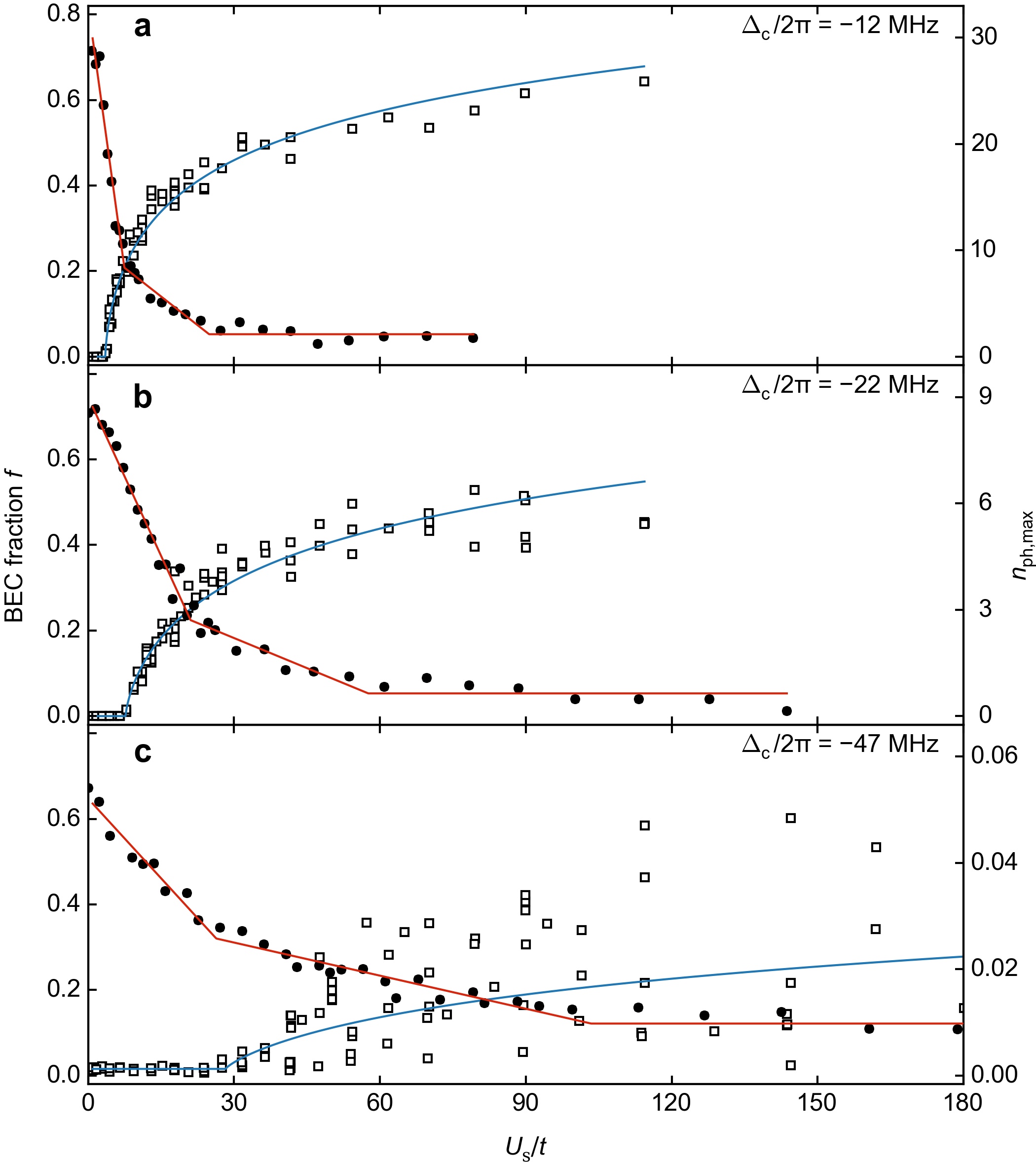}
\caption{{\bf Determination of the phase boundaries.} BEC fraction $f$ (averaged into 100 equally spaced bins) and maximum photon number $n_{\mathrm{ph,max}}$ (closed and open symbols, respectively) as a function of $U_{\mathrm{s}}/t$ for detunings $\Delta_\mathrm{c}/2\pi$ of $-12\, \unit{MHz}$ ({\bf a}), $-22\, \unit{MHz}$ ({\bf b}) and $-47\, \unit{MHz}$~({\bf c}). The red curve shows the result of a piecewise linear fit to $f$. We confirmed that the initial BEC fraction has no systematic dependence on $\Delta_{\mathrm{c}}$. The blue curve displays a power law fit to $n_{\mathrm{ph,max}}$. }
\label{fig:ext1}
\end{figure}
\subsection{Self-consistent checkerboard lattice.}
The SS and CDW phases give rise to a light field inside the cavity due to the Bragg scattering of $z$-lattice photons. This cavity field is self-consistent as it depends on the strength of the $\lambda$-periodic density modulation in the atomic cloud and has a depth $V_{\mathrm{c}}=n_{\mathrm{ph}} \times 12.3 \times 10^{-3} \,E_{\mathrm{R}}$. Interference of this self-consistent $x$-lattice with the field of the $z$-lattice produces a checkerboard lattice potential of depth $V_{\mathrm{CB}} = 2\sqrt{V_{\mathrm{2D}} V_{\mathrm{c}}}$, displayed in Extended Data Fig. 6. The line of constant $V_{\mathrm{CB}}$ bends towards smaller values of $V_{\mathrm{2D}}$ when approaching cavity resonance, substantiating the assumption that this energy offset causes the observed behaviour of the SS to CDW boundary line.
When the energy offset between even and odd sites due to $V_{\mathrm{CB}}$ gets comparable to the tunnelling energy, the effective tunnelling strength between nearest-neighbours reduces and higher order tunnelling processes begin to play a significant role \cite{Bissbort}.
\begin{figure}
\includegraphics[width=0.98\columnwidth]{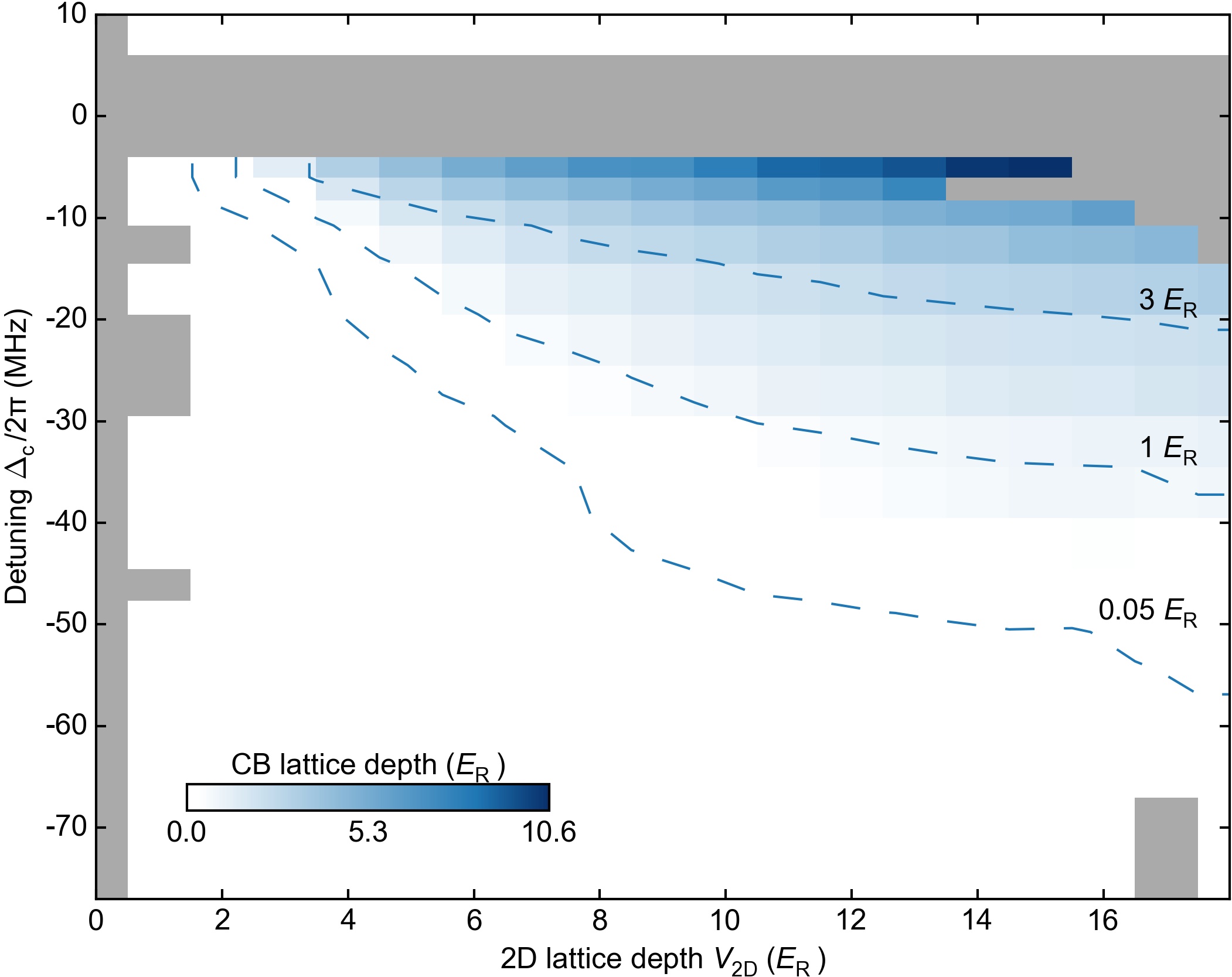}
\caption{{\bf Strength of self-consistent checkerboard (CB) lattice.} The CB lattice depth extracted from the measured mean intracavity photon number $n_{\mathrm{ph}}$, as a function of the applied lattice depth $V_{\mathrm{2D}}$ and detuning $\Delta_{\mathrm{c}}$. The CB lattice depth gets comparable to the depth of the static lattices close to cavity resonance, but drops rapidly when moving away due to its detuning dependence. Exemplary equipotential lines at $0.05\,E_{\mathrm{R}}$, $1\,E_{\mathrm{R}}$ and $3\,E_{\mathrm{R}}$ are shown.}
\label{fig:extCB}
\end{figure}

\subsection{Hysteresis measurements.}
We initialize the system in the insulating region at either $V_{\mathrm{2D}}= 14\,E_{\mathrm{R}}$ or $18\,E_{\mathrm{R}}$ using a $50\,\unit{ms}$ long S-shaped amplitude ramp. The detuning $\Delta_{\mathrm{c}}$ is then changed with an S-shaped frequency ramp at an average speed of $0.67 \,\unit{MHz/ms}$ reaching a different detuning value. After holding for $10\,\unit{ms}$, we scan back to the initial detuning. Residual atom loss continuously reduces the measured mean intracavity photon number $n_{\mathrm{ph}}$, which we take into account by rescaling the data before converting it into an imbalance $\Theta$. The scaling factor is extracted from reference measurements, where we hold at different $\Delta_{\mathrm{c}}$ for $50\,\unit{ms}$. We deduce a linear decrease in $n_{\mathrm{ph}}$ by $48(4)\%$ ($41(4)\%)$ for lattice depths of $V_{\mathrm{2D}}=14\,E_{\mathrm{R}}$ $(18 \,E_{\mathrm{R}}$). After rescaling the data, we observe a remaining relative drift of the imbalance level of $8(4)\%$ during the hold time.
Extended Data Fig. 7 shows detuning scans performed at $V_{\mathrm{2D}}=18\,E_{\mathrm{R}}$, where a similar hysteretic behaviour is observed as in Fig. 4 of the main text.  To test the sensitivity of the hysteretic behaviour on the ramp speed, we slow down the frequency ramp by a factor of two and observe a comparable evolution of the even-odd imbalance, see Extended Data Fig. 7. 
\begin{figure}
\includegraphics[width=0.98\columnwidth]{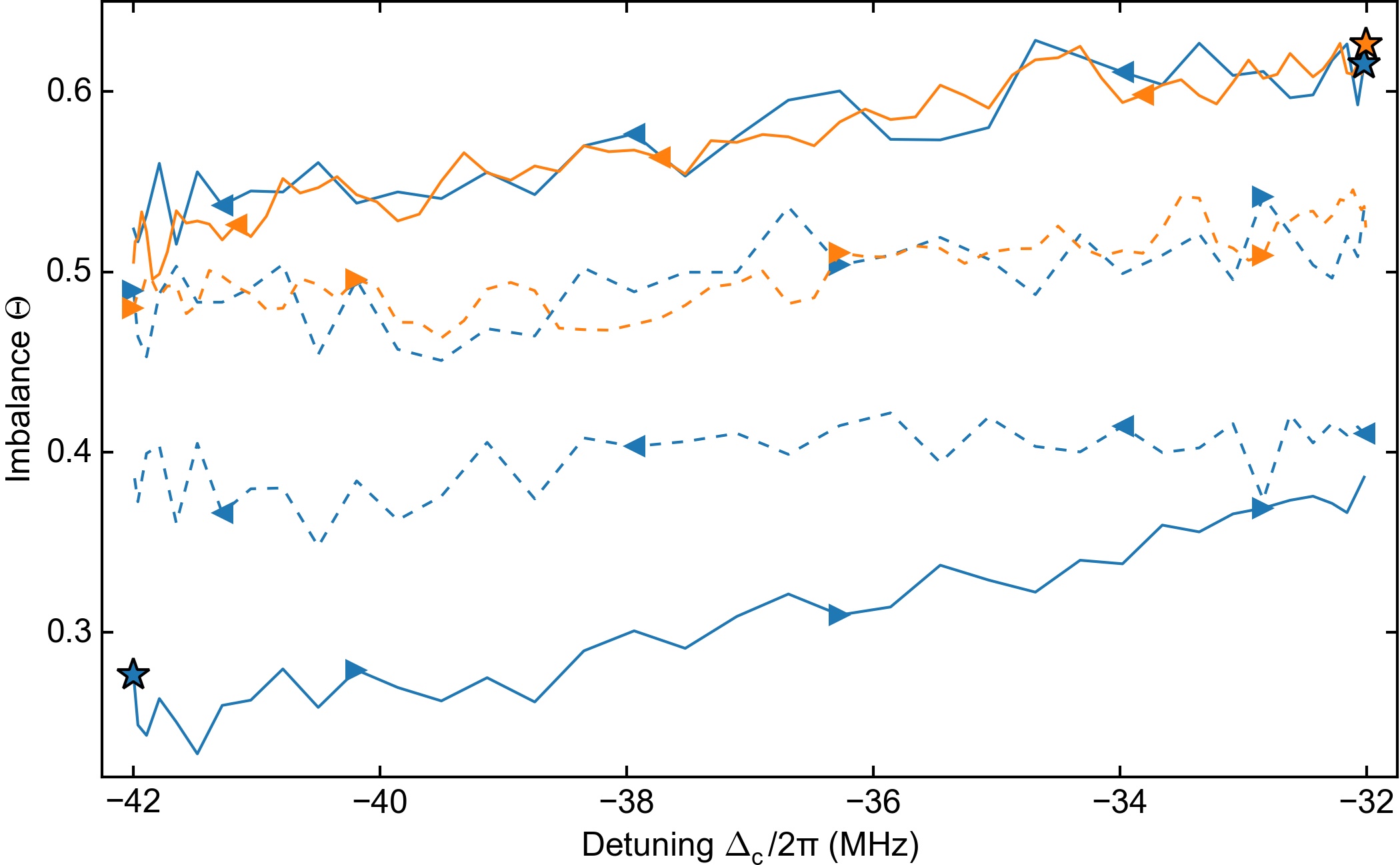}
\caption{{\bf Sensitivity on the ramp speed.} Hysteretic behaviour in the insulating regime, at $V_{\mathrm{2D}}=18\, E_{\mathrm{R}}$. The detuning $\Delta_{\mathrm{c}}$ is ramped at two speeds, $0.67 \, \unit{MHz/ms}$ (blue) and $0.33 \,\unit{MHz/ms}$ (orange). Lines result from an average of two to five measurements, using $400\,\unit{\mu s}$ time bins. Stars signify starting points, arrows show the scan direction and dashed lines indicate the return back to the starting point.}
\label{fig:extdetuning}
\end{figure}

\bibliographystyle{naturemag}

\begin{thebibliography}{10}
\expandafter\ifx\csname url\endcsname\relax
  \def\url#1{\texttt{#1}}\fi
\expandafter\ifx\csname urlprefix\endcsname\relax\def\urlprefix{URL }\fi
\providecommand{\bibinfo}[2]{#2}
\providecommand{\eprint}[2][]{\url{#2}}

\bibitem{Baier2015}
\bibinfo{author}{Baier, S.} \emph{et~al.}
\newblock \bibinfo{title}{{Extended Bose-Hubbard Models with Ultracold Magnetic
  Atoms}}.
\newblock \emph{\bibinfo{journal}{arXiv preprint}} \bibinfo{pages}{1507.03500}
  (\bibinfo{year}{2015}).

\bibitem{Yan2013}
\bibinfo{author}{Yan, B.} \emph{et~al.}
\newblock \bibinfo{title}{{Observation of dipolar spin-exchange interactions
  with lattice-confined polar molecules.}}
\newblock \emph{\bibinfo{journal}{Nature}} \textbf{\bibinfo{volume}{501}},
  \bibinfo{pages}{521--525} (\bibinfo{year}{2013}).

\bibitem{Baumann2010}
\bibinfo{author}{Baumann, K.}, \bibinfo{author}{Guerlin, C.},
  \bibinfo{author}{Brennecke, F.} \& \bibinfo{author}{Esslinger, T.}
\newblock \bibinfo{title}{{Dicke quantum phase transition with a superfluid gas
  in an optical cavity.}}
\newblock \emph{\bibinfo{journal}{Nature}} \textbf{\bibinfo{volume}{464}},
  \bibinfo{pages}{1301--1306} (\bibinfo{year}{2010}).

\bibitem{Mottl2012}
\bibinfo{author}{Mottl, R.} \emph{et~al.}
\newblock \bibinfo{title}{{Roton-Type Mode Softening in a Quantum Gas with
  Cavity-Mediated Long-Range Interactions}}.
\newblock \emph{\bibinfo{journal}{Science}} \textbf{\bibinfo{volume}{336}},
  \bibinfo{pages}{1570--1573} (\bibinfo{year}{2012}).

\bibitem{Bloch2012}
\bibinfo{author}{Bloch, I.}, \bibinfo{author}{Dalibard, J.} \&
  \bibinfo{author}{Nascimb\`{e}ne, S.}
\newblock \bibinfo{title}{{Quantum simulations with ultracold quantum gases}}.
\newblock \emph{\bibinfo{journal}{Nature Phys.}} \textbf{\bibinfo{volume}{8}},
  \bibinfo{pages}{267--276} (\bibinfo{year}{2012}).

\bibitem{Bloch2002}
\bibinfo{author}{Greiner, M.}, \bibinfo{author}{Mandel, O.},
  \bibinfo{author}{Esslinger, T.}, \bibinfo{author}{H\"{a}nsch, T.~W.} \&
  \bibinfo{author}{Bloch, I.}
\newblock \bibinfo{title}{{Quantum phase transition from a superfluid to a Mott
  insulator in a gas of ultracold atoms}}.
\newblock \emph{\bibinfo{journal}{Nature}} \textbf{\bibinfo{volume}{415}},
  \bibinfo{pages}{39--44} (\bibinfo{year}{2002}).

\bibitem{Kohl2005}
\bibinfo{author}{K\"{o}hl, M.}, \bibinfo{author}{Moritz, H.},
  \bibinfo{author}{St\"{o}ferle, T.}, \bibinfo{author}{Schori, C.} \&
  \bibinfo{author}{Esslinger, T.}
\newblock \bibinfo{title}{{Superfluid to Mott insulator transition in one, two,
  and three dimensions}}.
\newblock \emph{\bibinfo{journal}{J. Low Temp. Phys.}}
  \textbf{\bibinfo{volume}{138}}, \bibinfo{pages}{635--644}
  (\bibinfo{year}{2005}).

\bibitem{Weiner1989}
\bibinfo{author}{Weiner, J.}, \bibinfo{author}{Bagnato, V.~S.},
  \bibinfo{author}{Zilio, S.} \& \bibinfo{author}{Julienne, P.~S.}
\newblock \bibinfo{title}{{Experiments and theory in cold and ultracold
  collisions}}.
\newblock \emph{\bibinfo{journal}{Rev. Mod. Phys.}}
  \textbf{\bibinfo{volume}{71}}, \bibinfo{pages}{1--85} (\bibinfo{year}{1999}).

\bibitem{Ni2008}
\bibinfo{author}{Ni, K.-K.} \emph{et~al.}
\newblock \bibinfo{title}{{A High Phase-Space-Density Gas of Polar Molecules}}.
\newblock \emph{\bibinfo{journal}{Science}} \textbf{\bibinfo{volume}{322}},
  \bibinfo{pages}{231--235} (\bibinfo{year}{2008}).

\bibitem{Stuhler2005}
\bibinfo{author}{Stuhler, J.} \emph{et~al.}
\newblock \bibinfo{title}{{Observation of Dipole-Dipole Interaction in a
  Degenerate Quantum Gas}}.
\newblock \emph{\bibinfo{journal}{Phys. Rev. Lett.}}
  \textbf{\bibinfo{volume}{95}}, \bibinfo{pages}{150406}
  (\bibinfo{year}{2005}).

\bibitem{Heidemann2008}
\bibinfo{author}{Heidemann, R.} \emph{et~al.}
\newblock \bibinfo{title}{{Rydberg Excitation of Bose-Einstein Condensates}}.
\newblock \emph{\bibinfo{journal}{Phys. Rev. Lett.}}
  \textbf{\bibinfo{volume}{100}}, \bibinfo{pages}{033601}
  (\bibinfo{year}{2008}).


\bibitem{Dutta2015}
\bibinfo{author}{Dutta, O.} \emph{et~al.}
\newblock \bibinfo{title}{{Non-standard Hubbard models in optical lattices: a
  review}}.
\newblock \emph{\bibinfo{journal}{Rep. Prog. Phys.}}
  \textbf{\bibinfo{volume}{78}}, \bibinfo{pages}{066001}
  (\bibinfo{year}{2015}).

\bibitem{Mickiewicz1990}
\bibinfo{author}{Micnas, R.}, \bibinfo{author}{Ranninger, J.} \&
  \bibinfo{author}{Robaszkiewicz, S.}
\newblock \bibinfo{title}{{Superconductivity in narrow-band systems with local
  nonretarded attractive interactions}}.
\newblock \emph{\bibinfo{journal}{Rev. Mod. Phys.}}
  \textbf{\bibinfo{volume}{62}}, \bibinfo{pages}{113--171}
  (\bibinfo{year}{1990}).

\bibitem{Goral2002}
\bibinfo{author}{G\'{o}ral, K.}, \bibinfo{author}{Santos, L.} \&
  \bibinfo{author}{Lewenstein, M.}
\newblock \bibinfo{title}{{Quantum phases of dipolar bosons in optical
  lattices.}}
\newblock \emph{\bibinfo{journal}{Phys. Rev. Lett.}}
  \textbf{\bibinfo{volume}{88}}, \bibinfo{pages}{170406}
  (\bibinfo{year}{2002}).

\bibitem{Kovrizhin2004}
\bibinfo{author}{Kovrizhin, D.~L.}, \bibinfo{author}{Pai, G.~V.} \&
  \bibinfo{author}{Sinha, S.}
\newblock \bibinfo{title}{{Density wave and supersolid phases of correlated
  bosons in an optical lattice}}.
\newblock \emph{\bibinfo{journal}{EPL}} \textbf{\bibinfo{volume}{72}},
  \bibinfo{pages}{162--168} (\bibinfo{year}{2005}).

\bibitem{VanOtterlo1995}
\bibinfo{author}{{Van Otterlo}, A.} \emph{et~al.}
\newblock \bibinfo{title}{{Quantum phase transitions of interacting bosons and
  the supersolid phase}}.
\newblock \emph{\bibinfo{journal}{Phys. Rev. B}} \textbf{\bibinfo{volume}{52}},
  \bibinfo{pages}{16176--16186} (\bibinfo{year}{1995}).

\bibitem{Scarola2005}
\bibinfo{author}{Scarola, V.~W.} \& \bibinfo{author}{Sarma, S.~D.}
\newblock \bibinfo{title}{{Quantum Phases of the Extended Bose-Hubbard
  Hamiltonian: Possibility of a Supersolid State of Cold Atoms in Optical
  Lattices}}.
\newblock \emph{\bibinfo{journal}{Phys. Rev. Lett.}}
  \textbf{\bibinfo{volume}{95}}, \bibinfo{pages}{33003} (\bibinfo{year}{2005}).

\bibitem{DallaTorre2006}
\bibinfo{author}{{Dalla Torre}, E.~G.}, \bibinfo{author}{Berg, E.} \&
  \bibinfo{author}{Altman, E.}
\newblock \bibinfo{title}{{Hidden Order in 1D Bose Insulators}}.
\newblock \emph{\bibinfo{journal}{Phys. Rev. Lett.}}
  \textbf{\bibinfo{volume}{97}}, \bibinfo{pages}{260401}
  (\bibinfo{year}{2006}).
  
  \bibitem{Klinder2015a}
\bibinfo{author}{Klinder, J.} \emph{et~al.}
\newblock \bibinfo{title}{{Observation of a Superradiant Mott Insulator in the
  Dicke-Hubbard Model}}.
  \newblock \emph{\bibinfo{journal}{Phys. Rev. Lett.}}
  \textbf{\bibinfo{volume}{115}}, \bibinfo{pages}{230403}
  (\bibinfo{year}{2015}).

\bibitem{Ritsch2013}
\bibinfo{author}{Ritsch, H.}, \bibinfo{author}{Domokos, P.},
  \bibinfo{author}{Brennecke, F.} \& \bibinfo{author}{Esslinger, T.}
\newblock \bibinfo{title}{{Cold atoms in cavity-generated dynamical optical
  potentials}}.
\newblock \emph{\bibinfo{journal}{Rev. Mod. Phys.}}
  \textbf{\bibinfo{volume}{85}}, \bibinfo{pages}{553--601}
  (\bibinfo{year}{2013}).

\bibitem{Li2013}
\bibinfo{author}{Li, Y.}, \bibinfo{author}{He, L.} \&
  \bibinfo{author}{Hofstetter, W.}
\newblock \bibinfo{title}{{Lattice-supersolid phase of strongly correlated
  bosons in an optical cavity}}.
\newblock \emph{\bibinfo{journal}{Phys. Rev. A}} \textbf{\bibinfo{volume}{87}},
  \bibinfo{pages}{051604} (\bibinfo{year}{2013}).

\bibitem{Habibian2013a}
\bibinfo{author}{Habibian, H.}, \bibinfo{author}{Winter, A.},
  \bibinfo{author}{Paganelli, S.}, \bibinfo{author}{Rieger, H.} \&
  \bibinfo{author}{Morigi, G.}
\newblock \bibinfo{title}{{Bose-Glass Phases of Ultracold Atoms due to Cavity
  Backaction}}.
\newblock \emph{\bibinfo{journal}{Phys. Rev. Lett.}}
  \textbf{\bibinfo{volume}{110}}, \bibinfo{pages}{075304}
  (\bibinfo{year}{2013}).
  
\bibitem{Baumann2011}
\bibinfo{author}{Baumann, K.}, \bibinfo{author}{Mottl, R.},
  \bibinfo{author}{Brennecke, F.} \& \bibinfo{author}{Esslinger, T.}
\newblock \bibinfo{title}{{Exploring symmetry breaking at the Dicke quantum
  phase transition}}.
\newblock \emph{\bibinfo{journal}{Phys. Rev. Lett.}}
  \textbf{\bibinfo{volume}{107}}, \bibinfo{pages}{140402} 
  (\bibinfo{year}{2011}).

\bibitem{Jimenez-Garcia2010}
\bibinfo{author}{Jim\'{e}nez-Garc\'{\i}a, K.} \emph{et~al.}
\newblock \bibinfo{title}{{Phases of a Two-Dimensional Bose Gas in an Optical
  Lattice}}.
\newblock \emph{\bibinfo{journal}{Phys. Rev. Lett.}}
  \textbf{\bibinfo{volume}{105}}, \bibinfo{pages}{110401}
  (\bibinfo{year}{2010}).

\bibitem{Krauth2007}
\bibinfo{author}{Krauth, W.} \& \bibinfo{author}{Trivedi, N.}
\newblock \bibinfo{title}{{Mott and Superfluid Transitions in a Strongly
  Interacting Lattice Boson System}}.
\newblock \emph{\bibinfo{journal}{EPL}} \textbf{\bibinfo{volume}{14}},
  \bibinfo{pages}{627--632} (\bibinfo{year}{2007}).

\bibitem{Rigol2009}
\bibinfo{author}{Rigol, M.}, \bibinfo{author}{Batrouni, G.~G.},
  \bibinfo{author}{Rousseau, V.~G.} \& \bibinfo{author}{Scalettar, R.~T.}
\newblock \bibinfo{title}{{State diagrams for harmonically trapped bosons in
  optical lattices}}.
\newblock \emph{\bibinfo{journal}{Phys. Rev. A}} \textbf{\bibinfo{volume}{79}},
  \bibinfo{pages}{053605} (\bibinfo{year}{2009}).

\bibitem{Caballero-Benitez2015}
\bibinfo{author}{Caballero-Benitez, S.~F.} \& \bibinfo{author}{Mekhov, I.~B.}
\newblock \bibinfo{title}{{Quantum optical lattices for emergent many-body
  phases of ultracold atoms}}.
\newblock \emph{\bibinfo{journal}{Phys. Rev. Lett.}}
  \textbf{\bibinfo{volume}{115}}, \bibinfo{pages}{243604}
  (\bibinfo{year}{2015}).

\bibitem{Morsch2006}
\bibinfo{author}{Morsch, O.} \& \bibinfo{author}{Oberthaler, M.}
\newblock \bibinfo{title}{{Dynamics of Bose-Einstein condensates in optical
  lattices}}.
\newblock \emph{\bibinfo{journal}{Rev. Mod. Phys.}}
  \textbf{\bibinfo{volume}{78}}, \bibinfo{pages}{179--215}
  (\bibinfo{year}{2006}).

\bibitem{Stoferle2004}
\bibinfo{author}{St\"{o}ferle, T.}, \bibinfo{author}{Moritz, H.},
  \bibinfo{author}{Schori, C.}, \bibinfo{author}{K\"{o}hl, M.} \&
  \bibinfo{author}{Esslinger, T.}
\newblock \bibinfo{title}{{Transition from a Strongly Interacting 1D Superfluid
  to a Mott Insulator}}.
\newblock \emph{\bibinfo{journal}{Phys. Rev. Lett.}}
  \textbf{\bibinfo{volume}{92}}, \bibinfo{pages}{130403}
  (\bibinfo{year}{2004}).

\bibitem{Petrov2000}
\bibinfo{author}{Petrov, D.~S.}, \bibinfo{author}{Holzmann, M.} \&
  \bibinfo{author}{Shlyapnikov, G.~V.}
\newblock \bibinfo{title}{{Bose-Einstein Condensation in Quasi-2D Trapped
  Gases}}.
\newblock \emph{\bibinfo{journal}{Phys. Rev. Lett.}}
  \textbf{\bibinfo{volume}{84}}, \bibinfo{pages}{2551--2555}
  (\bibinfo{year}{2000}).

\bibitem{Maschler2007}
\bibinfo{author}{Maschler, C.}, \bibinfo{author}{Mekhov, I.~B.} \&
  \bibinfo{author}{Ritsch, H.}
\newblock \bibinfo{title}{{Ultracold atoms in optical lattices generated by
  quantized light fields}}.
\newblock \emph{\bibinfo{journal}{EPJ D}} \textbf{\bibinfo{volume}{46}},
  \bibinfo{pages}{545--560} (\bibinfo{year}{2008}).

\bibitem{Jaksch1998}
\bibinfo{author}{Jaksch, D.}, \bibinfo{author}{Bruder, C.},
  \bibinfo{author}{Cirac, J.~I.}, \bibinfo{author}{Gardiner, C.~W.} \&
  \bibinfo{author}{Zoller, P.}
\newblock \bibinfo{title}{{Cold Bosonic Atoms in Optical Lattices}}.
\newblock \emph{\bibinfo{journal}{Phys. Rev. A}} \textbf{\bibinfo{volume}{81}},
  \bibinfo{pages}{3108--3111} (\bibinfo{year}{1998}).

\bibitem{Spielman2008}
\bibinfo{author}{Spielman, I.}, \bibinfo{author}{Phillips, W.} \&
  \bibinfo{author}{Porto, J.}
\newblock \bibinfo{title}{{Condensate Fraction in a 2D Bose Gas Measured across
  the Mott-Insulator Transition}}.
\newblock \emph{\bibinfo{journal}{Phys. Rev. Lett.}}
  \textbf{\bibinfo{volume}{100}}, \bibinfo{pages}{120402}
  (\bibinfo{year}{2008}).

\bibitem{Landig2015}
\bibinfo{author}{Landig, R.}, \bibinfo{author}{Brennecke, F.},
  \bibinfo{author}{Mottl, R.}, \bibinfo{author}{Donner, T.} \&
  \bibinfo{author}{Esslinger, T.}
\newblock \bibinfo{title}{{Measuring the dynamic structure factor of a quantum
  gas undergoing a structural phase transition}}.
\newblock \emph{\bibinfo{journal}{Nat. Commun.}} \textbf{\bibinfo{volume}{6}},
  \bibinfo{pages}{7046} (\bibinfo{year}{2015}).

\bibitem{Bissbort}
\bibinfo{author}{Bissbort, U.}
\newblock \bibinfo{title}{{private communication}}.

\end{thebibliography}

\textbf{Acknowledgements} 
We thank U. Bissbort, G. Graf, S. Huber, G. Morigi, L. Pollet and H. Ritsch for discussions and F. Brennecke for contributions in the early design phase of the experiment. Financial funding from Synthetic Quantum Many-Body Systems (European Research Council advanced grant), the EU Collaborative Project TherMiQ (Grant Agreement 618074), SBFI support for Horizon2020 project QUIC and SNF support for the DACH project 'Quantum Crystals of Matter and Light'.

\textbf{Author Contributions} 
R.L., L.H., N.D. and M.L. took the data and analyzed them together with T.D. Contributions to the design of the experiment were made by R.M. All work was supervised by T.E. All authors contributed to discussions and the preparation of the manuscript. 

\textbf{Author Information} 
The authors declare no competing financial interests. Correspondence and requests for materials should be addressed to T.D. (donner@phys.ethz.ch).  
\end{document}